\documentclass[11pt, letterpaper, logo, onecolumn, copyright, numbering]{acc}

\usepackage{url}
\usepackage{natbib}

\usepackage{amsthm}
\usepackage{times}

\definecolor{darkblue}{RGB}{32, 64, 129}
\definecolor{darkgreen}{RGB}{0, 110, 85}
\definecolor{darkred}{RGB}{153, 0, 0}
\definecolor{graytext}{gray}{0.45}

\definecolor{evaluatorcolor}{RGB}{255,230,230}    %
\definecolor{evaluatorframe}{RGB}{180,50,50}      %
\definecolor{taskgencolor}{RGB}{230,255,230}      %
\definecolor{taskgenframe}{RGB}{50,180,50}        %
\definecolor{executioncolor}{RGB}{230,230,255}    %
\definecolor{executionframe}{RGB}{50,50,180}      %
\definecolor{lightgray}{RGB}{240,240,240}
\definecolor{darkgray}{RGB}{80,80,80}

\usepackage{pifont}
\usepackage{longtable}

\usepackage{makecell}

\usepackage{amsmath}
\usepackage{pifont}

\usepackage{booktabs}
\usepackage{multirow}
\usepackage{graphicx}
\usepackage{float}
\usepackage{caption}
\usepackage{subcaption}
\usepackage{xspace}

\usepackage[utf8]{inputenc}
\usepackage[T1]{fontenc}
\usepackage{xcolor}
\usepackage[most]{tcolorbox}
\usepackage{enumitem}
\usepackage{listings}

\definecolor{darkgray}{rgb}{0.3, 0.3, 0.3}
\definecolor{lightgray}{rgb}{0.95, 0.95, 0.95}
\definecolor{codegray}{rgb}{0.98, 0.98, 0.98}

\newtcolorbox{promptbox}[1]{
    colback=lightgray,
    colframe=darkgray,
    colbacktitle=darkgray,
    coltitle=white,
    boxrule=2pt,
    arc=0mm,
    left=10pt,
    right=10pt,
    top=10pt,
    bottom=10pt,
    fonttitle=\bfseries\large,
    title={#1},
    attach boxed title to top left={yshift=-2mm}
}

\usepackage{wrapfig}

\usepackage{algorithm}
\usepackage[noend]{algpseudocode}

\definecolor{deepred}{rgb}{0.631,0.102,0.102}
\definecolor{skyblue}{HTML}{126da2}

\definecolor{accpurple}{HTML}{A100FF}
\hypersetup{
     colorlinks = true,
     breaklinks = true,
     linkcolor = accpurple,
     urlcolor = accpurple,
     citecolor = accpurple
     }

\definecolor{orange}{rgb}{1,0.5,0}

\algnewcommand{\LineComment}[1]{\State \(\triangleright\) #1}

\title{From Requirements to Attack Trees: Grounded LLM Agents for Design-Time Security Review}

\author[1, *]{Akash Iyer}
  \author[1, *]{Taha Demirkan}
  \author[1]{Keerthi Koneru}
  \author[1]{Aaryan Siddharthan}
  \author[1]{Sheethal Kumar}
  \author[1]{Ramesh Radhakrishnan}

  \affil[1]{Center for Advanced AI, Accenture}

  \date{\today}
  \correspondingauthor{Keerthi Koneru (\href{mailto:keerthi.koneru@accenture.com}{keerthi.koneru@accenture.com}). $^*$These authors contributed equally to this work. \\This paper is an Accenture technical preprint.}

\begin{document}
\begin{abstract}
Design-level security weaknesses can arise from requirements, trust assumptions, missing controls, and data flows before implementation begins. 
Existing security practices often identify these issues after code is written. 
We present a multi-agent LLM framework for design-time security analysis from product requirement documents and architecture diagrams.

The proposed framework parses architecture diagrams into graph representations, generates misuse and failure cases, constructs attack trees, checks governance and compliance gaps, recommends mitigations, assigns enterprise security-domain tags, and produces a candidate revised architecture recommendation for expert review. 
The framework does not retrieve from Common Weakness Enumeration (CWE) databases at inference time. Instead, it analyzes system behavior, trust boundaries, component interactions, and data-flow assumptions. Misuse cases act as intermediate representations that link findings to system components and attack paths, while a validation and refinement loop filters unsupported findings and improves grounding, traceability, and actionability.

We evaluate the framework on a Microsoft reference-labeled threat-modeling example, labeled synthetic PRD--architecture pairs, and two open-ended systems: Berty and Gas Town. The reference-labeled case
supports threat-recovery and actionability analysis, while the open-ended
cases evaluate validity, noise, traceability, actionability, redundancy, and
attack-tree quality. Results show that architecture-informed, misuse-driven
reasoning improves review quality compared with single-shot and ablation
baselines.

\end{abstract}

\maketitle
\def\Snospace~{Section }
\def\sectionautorefname{\Snospace}
\def\subsectionautorefname{\Snospace}
\def\subsubsectionautorefname{\Snospace}
\def\chapterautorefname{\Snospace}

\section{Introduction}
Modern software systems are distributed, data-intensive, and deployed across
regulated domains. Security assurance often relies on post-implementation
techniques such as static analysis, dynamic testing, penetration testing, and
vulnerability scanning. These methods are useful for detecting implementation
defects, but they provide limited support for design-level weaknesses that arise
before code is written, including unsafe trust assumptions, missing controls,
insecure data flows, weak identity boundaries, and misplaced access paths.
Secure software-development and systems-security guidance emphasizes that
security should be integrated during requirements analysis and architecture
design~\cite{nist80053,ross2022secureSystems,souppaya2022ssdf,shevchenko2018threat}.
Prior work also shows that security weaknesses can originate from architecture
and design decisions and may not be exposed by code-level analysis
alone~\cite{houser2014designflaws}.

Threat modeling addresses this problem by analyzing how a system may be
misused, attacked, or fail under adversarial conditions. STRIDE (Spoofing, Tampering, Repudiation, Information Disclosure, Denial of Service, and Elevation of Privilege), misuse cases,
and attack trees provide structured ways to reason about threats, attacker
goals, and controls~\cite{shostack2014threat,schneier1999attack,
alexander2003misuse,sindre2005misuse}. However, practical threat modeling
remains difficult to scale because it requires security expertise, interpretation
of requirements, and architecture understanding. Enterprise review also spans
multiple teams, including IAM, network security, application security, data
security, cloud security, DevSecOps, and governance.

Recent LLM-assisted threat-modeling systems show that LLMs can generate
security-relevant findings from system descriptions~\cite{bandara2025astride,
jawad2025stridesea,elsharefThreatModelingLLM2024,raghavan2025stridegpt}.
However, direct prompting can produce unsupported findings, miss coverage gaps,
and generate generic threats that are not grounded in architecture. For design
review, findings must be traceable to requirements, components, data flows, trust
boundaries, attack paths, mitigations, and responsible security domains.

This paper presents a grounded multi-agent LLM framework for design-time
vulnerability discovery and secure architecture redesign from PRDs and
architecture diagrams. The framework parses diagrams into graph representations,
generates misuse and failure cases, constructs attack trees, identifies
governance gaps, recommends mitigations, assigns enterprise-domain tags, and
produces a revised architecture. Rather than retrieving from CVE databases at
inference time, it reasons over system behavior, trust assumptions, component
interactions, data flows, and control placement.

Misuse cases act as grounded intermediate representations connecting
requirements and architecture elements to adversarial goals, attack paths, and
mitigation actions. The framework uses iterative critique and revision
~\cite{madaan2023selfrefine} to validate and refine generated artifacts, reducing
unsupported findings while improving grounding, traceability, and actionability.

We evaluate the framework using a reference-labeled Microsoft threat-modeling
example and additional PRD--architecture case studies. The reference-labeled case
allows threat-recovery and actionability analysis, while the open-ended
cases evaluate validity, noise, traceability, actionability, redundancy, and
attack-tree quality.

We study three research questions:
\begin{itemize}
    \item \textbf{RQ1:} Can LLM agents identify design-time security weaknesses
    from PRDs and architecture diagrams?
    \item \textbf{RQ2:} Do misuse-case intermediates and validation
    improve grounding, traceability, and noise reduction over single-shot LLM
    analysis?
    \item \textbf{RQ3:} Can generated findings support mitigation planning,
    enterprise-domain routing, and secure architecture redesign?
\end{itemize}

\textbf{Contributions.}
This paper makes four contributions:
\begin{itemize}
    \item A grounded multi-agent LLM framework for design-time vulnerability
    discovery from PRDs and architecture diagrams.
    \item A misuse-case-centered workflow that links requirements, architecture
    elements, attack trees, governance findings, mitigations, and secure
    architecture redesign.
    \item Enterprise-domain tagging for routing findings across IAM, network,
    application, data, cloud, DevSecOps, and GRC teams.
    \item An evaluation using a reference-labeled threat-modeling example and
    ablation studies showing improved grounding, traceability, actionability,
    and noise reduction over single-shot baselines.
\end{itemize}
\section{Related Work}
Table~\ref{tab:comparison} summarizes representative prior approaches:
manual threat modeling~\cite{shostack2014threat,schneier1999attack,
alexander2003misuse,sindre2005misuse}, tool-based threat modeling~\cite{sion2020dfd,owaspThreatDragon,owaspPyTM,owaspAutomatedThreats}, enterprise threat modeling tools~\cite{threatmodeler2025}, security
design-flaw detection~\cite{tuma2020designFlaws}, architecture and DFD
extraction~\cite{schneider2023code2dfd,schneider2023microsecend}, and
LLM-assisted threat modeling~\cite{elsharefThreatModelingLLM2024,
raghavan2025stridegpt,bandara2025astride,jawad2025stridesea}, and highlights the functional capabilities of the proposed framework. 


{\textbf{Manual and Tool-Based Threat Modeling.}}
Foundational methods such as STRIDE (Spoofing, Tampering, Repudiation, Information Disclosure, Denial of Service, Elevation of Privilege), misuse cases, and attack trees provide
structured ways to reason about adversarial behavior, attacker goals, and
security controls~\cite{shostack2014threat,schneier1999attack,
alexander2003misuse,sindre2005misuse}. Misuse cases have also been used to
capture security requirements during design~\cite{johnstone2011misuse}. 
Existing tools improve documentation and repeatability but still depend on structured inputs or predefined threat patterns~\cite{sion2020dfd,owaspThreatDragon,owaspPyTM,threatmodeler2025}.

\begin{table*}[t]
\centering
\caption{\textbf{Comparison of Prior Approaches and Proposed Framework.}\\
\ding{51}: Full support. $\triangle$: Partial support — capability exists but requires
structured inputs, expert facilitation, or manual effort not automated by the approach
(e.g., Manual Threat Modeling supports Multi-Artifact Input partially because it can
consume heterogeneous artifacts but requires human interpretation rather than automated
parsing). \ding{55}: No support.}
\label{tab:comparison}
\resizebox{\linewidth}{!}{
\begin{tabular}{p{3.7cm}ccccccc}
\toprule
\textbf{Approach}
& \makecell{\textbf{Multi-Artifact}\\\textbf{Input}}
& \makecell{\textbf{Misuse}\\\textbf{Modeling}}
& \makecell{\textbf{Attack}\\\textbf{Trees}}
& \makecell{\textbf{Iterative}\\\textbf{Reasoning}}
& \makecell{\textbf{No CVE}\\\textbf{Retrieval}}
& \makecell{\textbf{Domain}\\\textbf{Routing}}
& \makecell{\textbf{Redesign}\\\textbf{Recommendation}} \\
\midrule

Manual Threat Modeling
& $\triangle$ & $\triangle$ & \ding{51} & \ding{55} & \ding{51} & $\triangle$ & \ding{55} \\

Tool-Based Threat Modeling
& $\triangle$ & \ding{55} & $\triangle$ & \ding{55} & $\triangle$ & \ding{55} & \ding{55} \\

Security Design Flaw Detection
& $\triangle$ & \ding{55} & \ding{55} & \ding{55} & \ding{51} & \ding{55} & \ding{55} \\

Architecture / DFD Extraction
& $\triangle$ & \ding{55} & \ding{55} & \ding{55} & \ding{51} & \ding{55} & \ding{55} \\

LLM Single-Prompt
& $\triangle$ & \ding{55} & \ding{55} & \ding{55} & $\triangle$ & \ding{55} & \ding{55} \\

LLM-Assisted Threat Modeling
& $\triangle$ & $\triangle$ & $\triangle$ & \ding{55} & $\triangle$ & \ding{55} & \ding{55} \\

\midrule
\textbf{Proposed Framework}
& \ding{51} & \ding{51} & \ding{51} & \ding{51} & \ding{51} & \ding{51} & \ding{51} \\

\bottomrule
\end{tabular}}
\end{table*}

{\textbf{Design-Flaw Detection and Architecture Extraction.}}
Prior work detects security design flaws from annotated design models and
extracts security-rich data-flow diagrams from software systems
~\cite{tuma2020designFlaws,schneider2023code2dfd,schneider2023microsecend}.
These approaches show the value of architecture-level security reasoning, but
they focus on structured models or extracted DFDs rather than end-to-end
reasoning over heterogeneous PRDs and architecture diagrams.

{\textbf{LLM-Assisted Threat Modeling and Refinement.}}
Recent LLM-based systems generate threat-modeling outputs from system
descriptions or retrieved security knowledge~\cite{elsharefThreatModelingLLM2024,
raghavan2025stridegpt,bandara2025astride,jawad2025stridesea}. These systems
demonstrate the feasibility of LLM-assisted security analysis, but many rely on
single-prompt, retrieval-based, or weakly structured inference. They do not use
misuse cases as validated intermediate representations for grounding attack
trees, governance findings, mitigations, and architecture redesign. Our work
adopts the general pattern of iterative critique and revision
~\cite{madaan2023selfrefine}, but applies it to structured security artifacts
grounded in architecture graphs, misuse-case mappings, attack-tree structure,
and mitigation consistency.

{\textbf{Research Gap.}}
Existing approaches either depend on manual expert-driven analysis, operate over
structured threat-modeling inputs, or generate LLM-based findings without
sufficient grounding and refinement. The proposed framework instead
combines multi-artifact input, misuse-case intermediates, planner--worker attack
tree construction, iterative validation, governance and mitigation reasoning,
enterprise-domain routing, and secure architecture redesign in a unified
LLM-agent pipeline.

\subsection{Key Differentiators of the Proposed Framework} Recent LLM-assisted threat-modeling approaches have demonstrated that large language models can identify security-relevant findings from system descriptions, requirements, and architectural artifacts. However, many existing approaches generate threats directly from prompts or retrieved security knowledge, often producing findings that are difficult to trace back to specific requirements, system components, trust boundaries, and data flows. Existing threat-modeling tools improve documentation and process repeatability, while architecture-analysis approaches focus on extracting structured representations from diagrams. Nevertheless, these approaches typically treat threat generation, validation, mitigation planning, and architecture review as separate activities rather than components of a unified security-analysis workflow 
\cite{shostack2014threat,schneier1999attack,
alexander2003misuse,sindre2005misuse}.

Tools such as stride-gpt and RAG-based security analyzers generate threats from natural-language system descriptions or retrieved vulnerability knowledge, but operate over unstructured text without parsing architecture graphs into formal representations. They do not use misuse cases as validated intermediate artifacts before attack-tree construction, and they do not perform iterative refinement to filter unsupported findings. The proposed framework differs in three ways: (1) it parses architecture diagrams into structured graph representations that ground every generated finding; (2) it uses misuse cases as a stable intermediate layer connecting requirements, components, attack paths, and mitigations before attack trees are constructed; and (3) it applies iterative critique and revision to validate and refine generated artifacts rather than producing single-pass outputs.

\paragraph{Misuse-case-centered reasoning.} Rather than treating threats as the primary output artifact, the framework introduces structured misuse and failure cases as intermediate security representations. These artifacts explicitly connect requirements, architecture elements, attack objectives, governance findings, mitigation recommendations, and architecture redesign actions. By maintaining these intermediate representations throughout the reasoning process, the framework improves traceability between identified weaknesses and supporting evidence while enabling downstream security analysis tasks \cite{alexander2003misuse,sindre2005misuse,johnstone2011misuse}. \paragraph{Architecture-grounded security analysis.} The framework transforms architecture diagrams into graph-based representations containing components, communication paths, trust boundaries, deployment zones, and data flows. Subsequent reasoning stages explicitly reference these architectural entities when generating misuse cases, attack trees, governance findings, and mitigation recommendations. This architecture-informed reasoning reduces reliance on generic threat generation and encourages findings that are grounded in the analyzed system structure rather than solely in textual descriptions \cite{schneider2023microsecend,schneider2023code2dfd}. \paragraph{Iterative validation and refinement.} Many existing LLM-assisted threat-modeling approaches rely on single-pass generation, where identified threats are reported directly as final outputs. In contrast, the proposed framework incorporates a validation and refinement stage that evaluates generated misuse cases and attack trees for relevance, feasibility, completeness, duplication, and architectural grounding. Artifacts classified as unsupported, incomplete, or overly broad are revised or regenerated through iterative feedback loops. This process is inspired by recent advances in iterative critique-and-revision reasoning and is designed to improve the consistency and actionability of generated security findings \cite{madaan2023selfrefine}. \paragraph{Integration with enterprise security-review workflows.} Beyond threat identification, the framework integrates governance assessment, mitigation generation, security-domain routing, and candidate architecture redesign recommendations within a single workflow. Findings are mapped to enterprise security domains such as IAM, Network Security, Application Security, Data Security, Cloud Security, DevSecOps, and GRC. This enables security findings to be routed to responsible review teams and supports human-in-the-loop security decision-making. The framework aims not only to identify potential design-time weaknesses but also to facilitate remediation planning and architecture improvement within enterprise software-development processes.
\section{Framework and Methodology}

\subsection{Objective}

The objective of this work is to support iterative security analysis
across software-development lifecycle checkpoints using evolving
requirements and architecture artifacts. The framework is designed
for enterprise security review, where approval is distributed across
multiple security domains. A generated finding must therefore be
grounded in the available artifacts, linked to affected requirements
or architecture elements, associated with a mitigation, and routed
to the team responsible for review.
\begin{figure*}[!t]
\centering
\includegraphics[width=\textwidth,height=0.33\textheight,keepaspectratio]{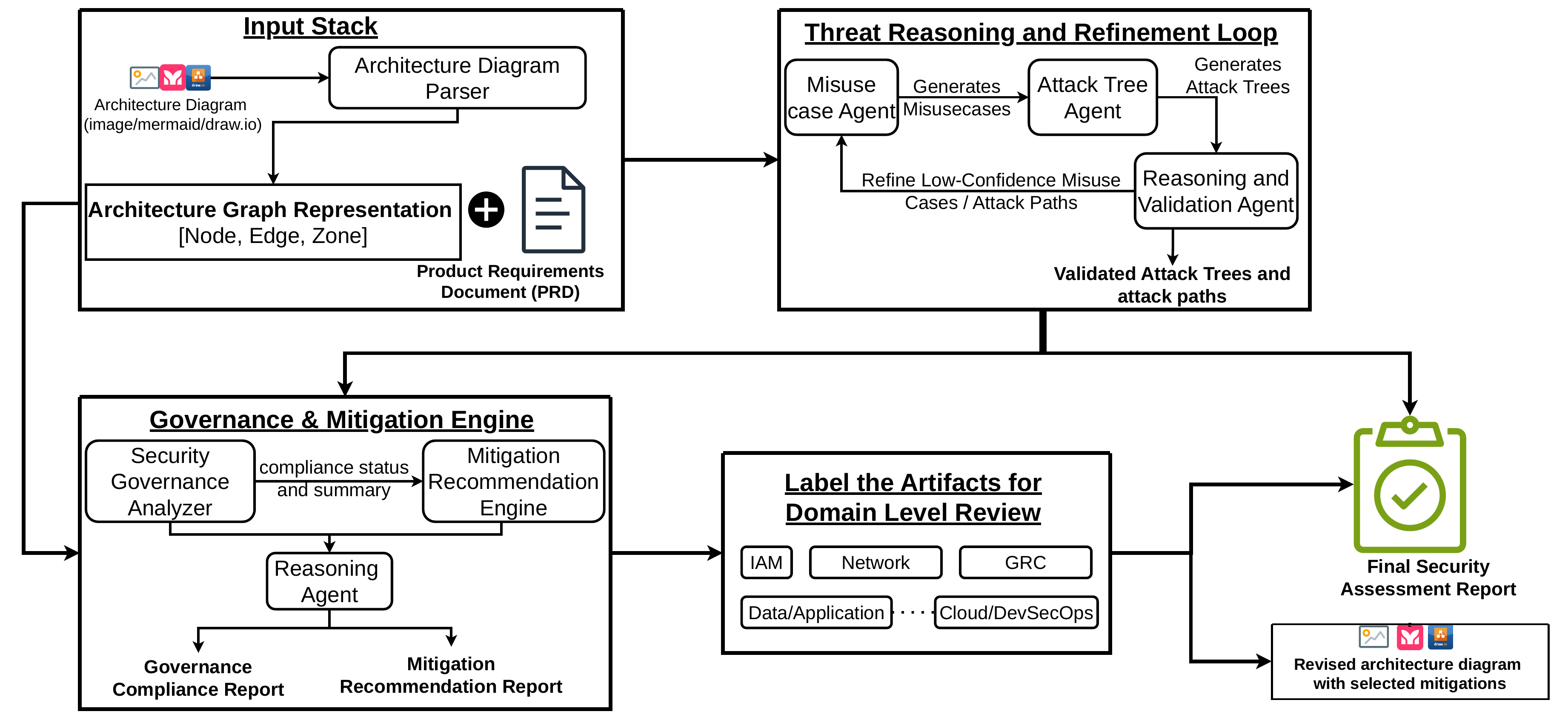}
\caption{Overview of the proposed multi-agent pipeline for iterative SDLC
security analysis from PRDs and architecture artifacts.}
\label{fig:architecture}
\end{figure*}

\subsection{Pipeline Overview}

Figure~\ref{fig:architecture} shows the proposed multi-agent security
reasoning pipeline. The pipeline has four stages: \textbf{input processing} (\ref{inputprocessing}),
\textbf{threat reasoning and refinement} (\ref{misusegen}, \ref{attackgen}, \ref{validation}), \textbf{governance and mitigation} (\ref{governance}), and
output representation and architecture redesign (\ref{output}).

\subsubsection{Requirement and Architecture Processing}
\label{inputprocessing}
The framework first normalizes the available design artifacts. From the PRD, it extracts functional requirements, actor roles, and integration dependencies, along with security expectations, compliance constraints, and data-handling assumptions.

When an architecture diagram is available, the framework converts it
into a graph representation $G=(N,E,Z)$. Nodes $N$ represent system
components such as services, APIs, databases, identity providers,
message queues, clients, and external systems. Edges $E$ represent
communication paths, data flows, protocols, and dependencies. Zones
$Z$ represent trust boundaries, deployment regions, cloud environments,
or logical system groups.

After parsing, each node and edge is annotated with its type, communication protocol, and data sensitivity, with trust-boundary crossings marked explicitly for downstream reasoning. The final graph is serialized
into a normalized textual representation for downstream agents.

\subsubsection{Misuse and Failure Case Generation}
\label{misusegen}
The misuse-case agent generates structured misuse and failure cases
from the PRD and from the architecture graph. A \textit{misuse case}
captures intentional adversarial behavior, while a \textit{failure case} captures
an accidental or operational condition that may lead to a security impact.

Each case is a structured record that captures the affected requirement or architecture element, the relevant actor and their entry point, the violated security property, and the conditions under which the case applies. It also includes an impact assessment, supporting evidence from the PRD or architecture graph, a confidence score, and a security-domain tag for routing. In architecture-informed mode, each case must be linked to one or more
nodes, edges, or zones in the architecture graph. In PRD-only mode, it
is linked to the supporting requirement text.

The misuse-case agent is designed before the attack-tree stage because misuse cases serve as grounding anchors. Generating attack trees directly from PRD and architecture context — without an intermediate validated representation — produces outputs that are difficult to trace back to specific components or requirements. By requiring each misuse case to be linked to architecture elements and scored for confidence before downstream use, the agent creates a stable unit of analysis that the attack-tree stage can expand systematically. This ordering also enables targeted regeneration: when a misuse case is revised or rejected during validation, only the attack-tree branches derived from it need to be regenerated, rather than rerunning the full pipeline.

The agent is instructed to reason across all architecture components and trust boundaries, considering diverse actor types and the full range of security properties  including authentication, authorization, data confidentiality and integrity, logging, and compliance obligations.

\subsubsection{Planner--Worker Attack Tree Construction}
\label{attackgen}
The attack-tree stage uses a planner--worker setup. The planner derives
a bounded set of attack objectives from validated misuse cases. Each
objective becomes the root of an attack tree. Worker agents then expand
each objective into attack paths using the associated misuse cases, PRD
context, and architecture graph. The planner–worker split is chosen to bound the complexity of attack-tree construction. A single agent generating all attack trees jointly risks producing flat or redundant trees because the generation context becomes dominated by earlier objectives. Separating planning (objective derivation) from expansion (path construction) allows each worker to focus on one attacker goal with its full misuse-case and architecture context, producing deeper and more coherent trees. The planner also applies a bound on the number of objectives to prevent over-generation when the misuse-case set is large.

Attack trees use three node types: AND nodes require all child steps to occur, OR nodes represent alternative attack paths, and LEAF nodes capture concrete actions, preconditions, or control weaknesses that an attacker can directly exploit. Each attack-tree node is linked back to supporting misuse
cases and, when available, architecture elements. This structure turns
isolated findings into traceable adversarial paths.

\subsubsection{Validation and Iterative Refinement}
\label{validation}
The validation agent evaluates misuse cases and attack trees.
Misuse cases are scored for relevance, likelihood, severity, and
confidence. Attack trees are scored for logical consistency, feasibility
of attack paths, completeness of preconditions, and grounding.

Each artifact is assigned one of five statuses: accepted, incomplete,
unsupported, duplicate, or needs-split. The actions are:
\begin{itemize}
    \item Remove or Regenerate unsupported artifacts.
    \item Revise incomplete artifacts.
    \item Merge duplicate artifacts.
    \item Broad findings are marked as needs-split and divided into more
specific cases.
\end{itemize}

The refinement loop runs for up to three iterations or until convergence. The validation agent is designed to operate on both misuse cases and attack trees jointly because the two artifact types are interdependent. A weak misuse case produces weak attack-tree branches, and an incoherent attack-tree branch often reveals that its originating misuse case lacked sufficient grounding. Evaluating them together in each iteration allows the loop to propagate corrections in both directions: low-confidence misuse cases trigger branch regeneration, and structurally weak branches trigger misuse-case revision. The three-iteration cap balances refinement quality against token cost; in practice, most runs converge in one or two iterations when the initial misuse-case generation is architecture-grounded.
Low-confidence misuse cases trigger the regeneration of related attack-tree
branches, and weak attack-tree branches trigger the refinement of the
originating misuse cases. Only affected artifacts are regenerated, and
artifact identifiers are preserved across iterations to maintain
traceability.

\subsubsection{Governance, Mitigation, and Domain Tagging}
\label{governance}
The governance and mitigation stage follows a generate--critique--filter
pattern. The governance analyzer identifies policy, compliance, and
control gaps from the validated misuse cases, attack trees, PRD context,
and architecture graph. Each governance finding includes the affected
artifact, violated control intent, severity, rationale, and remediation
guidance. 

The mitigation engine generates design-level remediation actions for
validated findings. Each mitigation includes affected artifacts,
recommended controls, implementation steps, rationale, priority, and expected
security effects. 
Mitigations are scoped to design-level architectural controls rather than implementation-level code changes. Priority is assigned based on the severity of the linked finding and the breadth of affected architecture components: findings that span multiple trust boundaries or affect high-sensitivity data flows are prioritized higher. The reasoning agent critiques mitigation recommendations for feasibility and specificity before inclusion in the final report, filtering out generic recommendations that are not traceable to the analyzed architecture.
Examples include policy enforcement points, enforcing
service-to-service authentication, 
separating trust zones, etc. 

Each generated artifact is assigned one or more enterprise security-domain
tags: IAM, Network Security, Application Security, Data Security, Cloud
Security, DevSecOps/Infrastructure Security, and GRC. These tags are applied
to misuse cases, attack-tree nodes, governance findings, mitigations, and
architecture changes so that findings can be routed to the responsible
review teams.

\subsubsection{Output Representation and Architecture Redesign}
\label{output}
The final output is a structured security assessment report containing
validated misuse and failure cases, attack trees, governance findings,
mitigation recommendations, domain tags, and reasoning traces. Misuse cases
and mitigations are represented in tabular form, while attack trees are
represented hierarchically.

The framework also generates a candidate revised architecture recommendation that incorporates selected mitigations. The revised design may add controls, components, edges,
or trust-boundary changes, such as an authorization service, policy
enforcement point, gateway, monitoring component, or isolated trust zone.
The revised architecture is 
only treated as a
structured recommendation for human security review and architecture
decision-making.

\section{Experimental Setup}

This section describes the evaluation artifacts, pipeline configurations, and
metrics. We use two types of evaluation inputs: (i) labeled evaluation cases — a reference threat-modeling example with existing threats and mitigations, and synthetic architecture--PRD pairs with manually reviewed ground-truth threats, and (ii) additional architecture/PRD
case studies without complete ground-truth threat annotations.

\subsection{Input Artifacts}
\label{dataset}

We are not aware of a public benchmark that provides paired PRDs, architecture diagrams, and complete ground-truth annotations for end-to-end design-time vulnerability
discovery and secure architecture redesign. We therefore evaluate the framework
using three types of artifacts: controlled flawed synthetic designs,
a reference-labeled threat-modeling example, and open-ended real-world
architecture/PRD case studies.

\textbf{Controlled flawed synthetic artifacts.}
We first evaluate the framework on intentionally flawed synthetic PRD--architecture
pairs. These artifacts were created to test whether the framework can identify
known design-time weaknesses under controlled conditions. Each PRD defines
functional requirements, security constraints, compliance expectations,
integrations, and data-handling assumptions. The corresponding Mermaid diagrams
include components, data flows, trust boundaries, and external systems.

We intentionally injected security gaps into the synthetic designs, including
explicit weaknesses such as missing authentication, and weak
service boundaries, as well as implicit omissions such as absent validation,
monitoring, policy enforcement, or control stages. These controlled flaws allow
us to evaluate direct detection, multi-step reasoning, mitigation generation, and
architecture redesign. Because the synthetic artifacts are purposefully flawed
and their ground truth is known, results on these cases are higher than on the
open-ended real-world systems.

For each synthetic case, an enterprise security expert and an enterprise
architect reviewed the PRD, architecture diagram, injected gaps, and expected
findings to produce the ground-truth threat set. Each ground-truth entry records
the threat, affected components, exploited weaknesses, and expected mitigation.
We categorize labels into misuse cases, which capture adversarial behavior, and workflow threats, which show architectural deficiencies or missing
controls.

\textbf{Reference-labeled threat-modeling example.}
We evaluate the framework on the Microsoft Code With Engineering Playbook
threat-modeling example~\cite{microsoftThreatModelExample}. This artifact
describes an IoT Edge video-processing application that runs modules on an edge
device and interacts with Azure services. It includes architecture and data-flow
diagrams, assets, trust assumptions, existing threats, and recommended
mitigations. We use it as a reference-labeled case because its existing threat
model allows measurement of threat recovery, actionability, and novelty
beyond the provided reference threats.

\textbf{Open-ended real-world case studies.}
We also evaluate the framework on Berty~\cite{berty2026} and Gas
Town~\cite{gastown2026}. Berty is an open-source secure peer-to-peer
messaging system, making it useful for evaluating design reasoning over
identity, privacy, networking, synchronization, and trust-boundary concerns. Gas
Town is an open-source multi-agent workspace manager, which lets us evaluate risks around agent coordination, tool access, shared state, workflow
execution, and authorization boundaries.

For each system, we construct a PRD from the README, architecture
documentation, and repository context. These cases assess whether the framework
can reason over real-world architectural complexity when no complete
ground-truth threat set is available. For these cases, we report validity, noise,
traceability, actionability, redundancy, and attack-tree quality rather than
coverage or precision against a gold threat set.

Additional details on the synthetic artifacts and scoring process are provided
in Appendix~\ref{appendix:synthetic-arch}.

\subsection{Pipeline Configurations}

We compare the full framework against two single-shot baselines and three
ablated configurations.

\begin{table}[!t]
\centering
\caption{Evaluation configurations.}
\label{tab:configs}
\small
\begin{tabular}{p{4.7cm}p{9.9cm}}
\toprule
\textbf{Configuration} & \textbf{Description} \\
\midrule
Single-shot LLM & One prompt generates misuse cases, attack trees, and mitigations from PRD and architecture. \\
Single-shot STRIDE & One prompt generates a STRIDE-structured threat table from PRD and architecture. \\
No-MC & Removes the misuse-case agent and generates attack trees directly from architecture context. \\
PRD-only & Uses only the PRD, evaluating early-stage security analysis without architecture grounding. \\
Arch-Informed & Full pipeline using PRD, architecture diagram, misuse cases, attack trees, and iterative refinement. \\
\bottomrule
\end{tabular}
\end{table}

All configurations use the same evaluation metrics, post-processing procedure,
LLM deployment, and decoding settings. We use \texttt{GPT-5 Mini} for all
agents, ablations, and baselines, with temperature set to 0 and a fixed seed of
42. We do not explicitly set \texttt{top\_p} or per-agent
\texttt{max\_tokens}, so provider defaults are used, except for the governance
agent, which is invoked through \texttt{get\_llm()} with \\
\texttt{max\_tokens=16000}. Each run uses static prompt templates and structured
Pydantic outputs where applicable, ensuring that prompt wording and output
schemas remain fixed across configurations.

No configuration retrieves from external CWE, CVE, or weakness databases during
inference. Governance findings are inferred from input artifacts and generated security artifacts, unless an explicit governance document is supplied
as part of the input. Each configuration is executed three times per system, and
reported metrics are averaged across the three runs.

Across runs, metric variation was low for the Arch-Informed configuration: on the synthetic cases, standard deviations for coverage, validity, and actionability were all below 0.03, indicating stable outputs under fixed temperature and seed. Variation was slightly higher for single-shot configurations, particularly on traceability and actionability, where the absence of structured intermediate artifacts makes outputs more sensitive to prompt sampling. 


Single-shot STRIDE is included as a structured baseline. Because it is prompted
to enumerate threats across STRIDE categories and relevant system domains, it can
generate overlapping findings. We therefore interpret its higher redundancy as a
trade-off of taxonomy-driven coverage rather than as distinct grounded threat
discovery.

The refinement loop runs for at most three feedback iterations and exits early
when no unsupported, incomplete, duplicate, or low-confidence artifacts require
revision.

\subsection{Evaluation Metrics}
\label{metrics}

We use two classes of metrics. For reference-labeled and synthetic cases with
manually reviewed ground truth, we report coverage, precision, and novelty metrics. For
all cases, including open-ended case studies without complete ground truth, we
report review-quality metrics such as validity, noise, traceability,
actionability, novelty, and attack-tree quality.

Let $T_{gt}$ denote the set of ground-truth threats, $T_{gen}$ the set of
generated threats, $T_{valid} \subseteq T_{gen}$ the validated threats,
$T_{mapped} \subseteq T_{valid}$ the threats grounded in the system architecture,
$T_{trace} \subseteq T_{valid}$ the threats linked to misuse cases or
intermediate reasoning artifacts, $T_{act} \subseteq T_{valid}$ the threats with
actionable mitigations, and $T_{dep} \subseteq T_{valid}$ the threats
incorporated into downstream reasoning artifacts such as attack trees.

\begin{table}[!t]
\centering
\caption{Evaluation metrics.}
\label{tab:metrics}
\small
\renewcommand{\arraystretch}{2.15}
\begin{tabular}{p{4.7cm}p{9.9cm}}
\toprule
\textbf{Metric} & \textbf{Definition} \\
\midrule

Coverage (Raw) &
Fraction of ground-truth threats identified by the system:
$\frac{|T_{gen} \cap T_{gt}|}{|T_{gt}|}$. \\

Grounded Coverage &
Fraction of ground-truth threats correctly identified and grounded in the
architecture:
$\frac{|T_{mapped} \cap T_{gt}|}{|T_{gt}|}$. \\

Grounded Precision &
Fraction of grounded generated threats that are correct:
$\frac{|T_{mapped} \cap T_{gt}|}{|T_{mapped}|}$. \\

Novelty &
Fraction of validated threats not present in the ground truth but judged
plausible:
$\frac{|T_{valid} \setminus T_{gt}|}{|T_{valid}|}$. \\

Validity &
Fraction of generated threats judged valid after evaluation:
$\frac{|T_{valid}|}{|T_{gen}|}$. \\

Noise &
Fraction of generated threats judged invalid or unsupported:
$1-\text{Validity}$. \\

Traceability &
Fraction of validated threats linked to misuse cases, architecture elements, or
intermediate reasoning artifacts:
$\frac{|T_{trace}|}{|T_{valid}|}$. \\

Actionability &
Fraction of validated threats with sensible mitigation recommendations:
$\frac{|T_{act}|}{|T_{valid}|}$. \\

Attack-Tree Quality &
Expert rating from 1--5 measuring attack-tree structure, concrete leaf actions,
sensible preconditions, and breadth of threat coverage. \\

\bottomrule
\end{tabular}
\end{table}

Coverage, grounded coverage, grounded precision, and novelty are reported only
for reference-labeled and synthetic cases where ground-truth threats are
available. For open-ended case studies without complete ground truth, these
metrics are not reported.

Generated threats were reviewed by an enterprise security expert using a fixed
rubric. A generated threat was counted as matching a reference threat when it
captured the same affected asset, weakness, and security objective, even if the
wording differed. Near-duplicate findings were merged before computing
redundancy. Novel findings were counted only when they were not present in the
reference set but were judged plausible for the analyzed architecture.

Attack-tree quality is scored on a 1--5 scale. A score of 1 indicates flat,
generic, or incoherent trees. A score of 3 indicates partially structured trees
with some valid decomposition and concrete leaf actions. A score of 5 indicates
multi-level trees with explicit AND/OR/LEAF structure, concrete leaf actions,
sensible preconditions, and broad coverage of the generated threat set.

Attack-tree quality was scored by a single enterprise security expert using the rubric above applied consistently across all configurations and systems. Because this is a subjective metric evaluated by one rater, inter-rater reliability was not formally assessed. This is acknowledged as a limitation in Section 7. To reduce within-rater inconsistency, all attack trees for a given system were scored in a single session in random configuration order, with the evaluator blinded to the configuration label.
\section{Results and Ablation Study}

We organize the results across three evaluation settings. First, we evaluate the
framework on controlled synthetic PRD--architecture pairs (referenced in Appendix~\ref{appendix:synthetic-arch}) with intentionally
injected design weaknesses and manually reviewed ground truth. These cases test
whether the framework can recover known design flaws under controlled
conditions, but their scores should be interpreted as controlled detection
results rather than evidence of broad real-world generalization. Second, we
evaluate on the Microsoft reference-labeled threat-modeling example
~\cite{microsoftThreatModelExample}, which includes existing threats and
mitigations. Third, we evaluate on Berty~\cite{berty2026} and Gas
Town~\cite{gastown2026}, two open-ended systems without complete
ground-truth threat annotations.

Across all evaluation settings, the Arch-Informed configuration provides the
most consistent performance. It does not always maximize a single metric, such
as raw coverage or number of generated findings, but it achieves the best overall
balance across recovery, grounding, traceability, actionability, and attack-tree
quality. Unlike these, some ablations improve isolated metrics by generating more
or less constrained outputs, but these gains often come with lower grounded
precision, weaker traceability, or higher redundancy.

\subsection{Controlled Synthetic Results}

The controlled synthetic artifacts include intentionally flawed Electronic Health System (EHS) and Online Banking System (OLB). The synthetic architectures and corresponding injected weaknesses are detailed in Appendix~\ref{appendix:synthetic-arch}. 
Because the ground truth is known and the designs are
purposefully flawed, these results are useful for controlled comparison but may
overestimate some evaluation metrics compared with open-ended real-world settings.

Table~\ref{tab:synthetic-results} reports the average results across three runs
for the two synthetic architectures. All
metrics except counts and attack-tree quality are reported as proportions. The
Arch-Informed configuration achieves the strongest overall balance. On EHS, it
achieves the highest raw coverage, grounded coverage, grounded precision,
traceability, actionability, and attack-tree quality. On OLB, No-MC achieves
slightly higher raw coverage, but its grounded precision is lower than
Arch-Informed, showing the cost of removing misuse-case intermediates.

\begin{table*}[t]
\centering
\caption{Controlled synthetic results averaged over three runs. Synthetic designs
contain intentionally injected weaknesses and manually reviewed ground truth.
For synthetic No-MC, coverage is computed over threats extracted from generated
attack-tree leaves; traceability is N/A because no misuse-case artifacts are
generated. \#MC: Misuse Cases Generated; \#AT: Attack Trees Generated.}
\label{tab:synthetic-results}
\scriptsize
\setlength{\tabcolsep}{2.5pt}
\renewcommand{\arraystretch}{1.12}
\resizebox{\textwidth}{!}{
\begin{tabular}{llrrrrrrrrrrrr}
\toprule
\textbf{Domain} & \textbf{Config.} &
\textbf{\#MC} & \textbf{\#AT} &
\textbf{Raw Cov.} & \textbf{G. Cov.} & \textbf{G. Prec.} &
\textbf{Valid.} & \textbf{Noise} & \textbf{Trace.} &
\textbf{Action.} & \textbf{Novel.} & \textbf{Redun.} &
\textbf{ATQ} \\
\midrule

EHS & Arch-Informed & 29.7 & 9.0 & 0.97 & 0.97 & 0.84 & 0.97 & 0.03 & 1.00 & 0.91 & 0.63 & 0.13 & 4.40 \\
EHS & No-MC & 0.0 & 15.3 & 0.67 & 0.67 & 0.26 & 0.99 & 0.01 & N/A & 0.20 & 0.73 & 0.13 & 4.30 \\
EHS & PRD-only & 30.3 & 11.7 & 0.88 & 0.85 & 0.72 & 0.95 & 0.05 & 0.72 & 0.88 & 0.72 & 0.06 & 4.13 \\
EHS & Single-shot LLM & 20.7 & 4.3 & 0.44 & 0.40 & 0.24 & 0.88 & 0.12 & 0.00 & 0.06 & 0.66 & 0.23 & 3.45 \\
EHS & Single-shot STRIDE & 23.3 & 4.0 & 0.58 & 0.42 & 0.36 & 0.87 & 0.13 & 0.00 & 0.31 & 0.56 & 0.17 & 3.30 \\

\midrule

OLB & Arch-Informed & 11.3 & 13.0 & 0.85 & 0.85 & 0.81 & 0.99 & 0.01 & 0.80 & 0.90 & 0.25 & 0.10 & 4.30 \\
OLB & No-MC & 0.0 & 17.3 & 0.86 & 0.86 & 0.68 & 0.95 & 0.05 & N/A & 0.92 & 0.12 & 0.07 & 4.17 \\
OLB & PRD-only & 15.0 & 12.7 & 0.83 & 0.82 & 0.72 & 0.95 & 0.05 & 0.82 & 0.90 & 0.33 & 0.18 & 4.20 \\
OLB & Single-shot LLM & 21.0 & 5.7 & 0.53 & 0.49 & 0.35 & 0.83 & 0.17 & 0.00 & 0.11 & 0.19 & 0.11 & 3.33 \\
OLB & Single-shot STRIDE & 21.7 & 4.3 & 0.58 & 0.50 & 0.43 & 0.85 & 0.15 & 0.00 & 0.30 & 0.50 & 0.15 & 3.50 \\

\bottomrule
\end{tabular}}
\end{table*}

On EHS, Arch-Informed reaches 0.97 raw and grounded coverage, 0.84 grounded
precision, 1.00 traceability, 0.91 actionability, and 4.40 attack-tree quality.
The single-shot LLM baseline performs substantially worse, with 0.44 raw
coverage, 0.24 grounded precision, 0.00 traceability, and 0.06 actionability.
Single-shot STRIDE improves over the single-shot LLM baseline in raw coverage
(0.58 vs. 0.44) and grounded precision (0.36 vs. 0.24), but does not match Arch-Informed across grounded coverage, traceability, actionability, and
attack-tree quality. Its higher redundancy reflects overlapping findings caused
by STRIDE-category enumeration.

On OLB, No-MC reaches slightly higher raw and grounded coverage than
Arch-Informed, but its grounded precision drops to 0.68, compared with 0.81 for
Arch-Informed. This shows that removing misuse-case intermediates can increase
apparent recall while reducing grounded correctness. Single-shot STRIDE again
improves over the single-shot LLM baseline across raw coverage, grounded
coverage, grounded precision, actionability, and attack-tree quality, but it
still trails Arch-Informed on the metrics most important for review quality.
Overall, Arch-Informed provides the best balance across grounded precision,
validity, low noise, traceability, actionability, and attack-tree quality.

\subsection{Reference-Labeled Microsoft Threat Model}

\begin{figure}[t]
\centering
\includegraphics[width=0.75\linewidth]{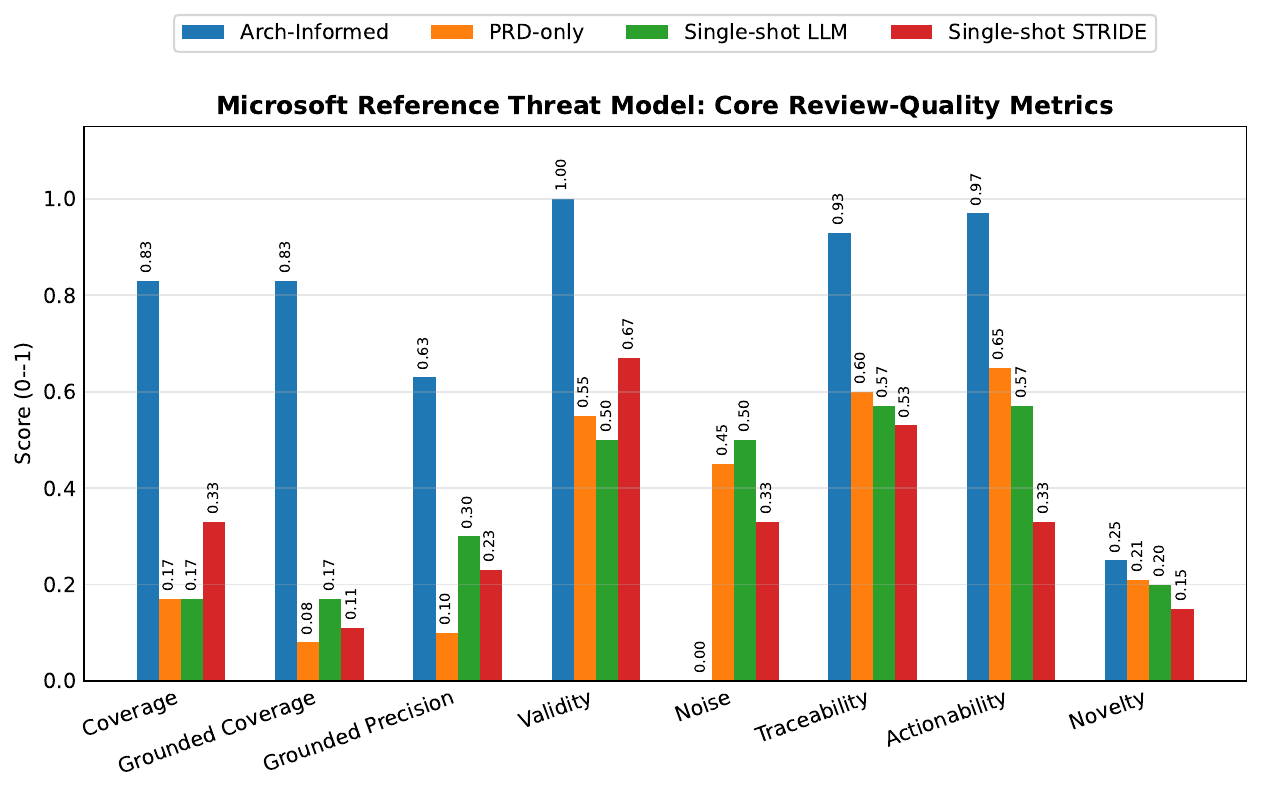}
\caption{Core metrics on the Microsoft reference-labeled threat-modeling
example. No-MC is excluded because threat-scoped metrics are not applicable
when the misuse-case agent is removed.}
\label{fig:microsoft-core}
\end{figure}

\begin{figure}[!t]
\centering
\includegraphics[width=0.75\linewidth]{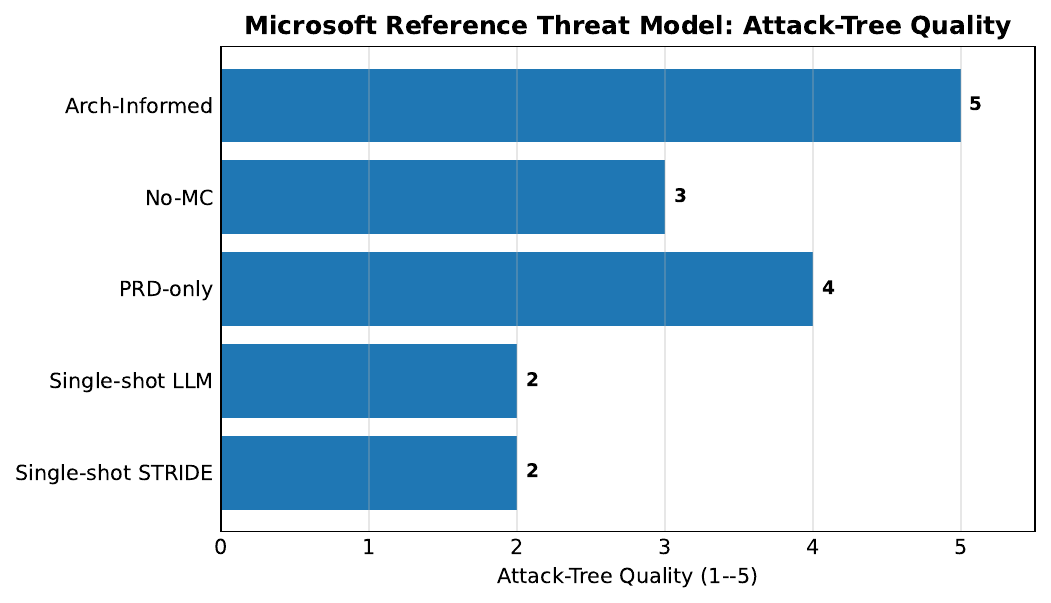}
\caption{Attack-tree quality on the Microsoft reference-labeled
threat-modeling example. No-MC is included because it still produces downstream
attack-tree artifacts.}
\label{fig:microsoft-atq}
\end{figure}

The Microsoft threat-modeling example includes an existing threat list and recommended
mitigations, allowing threat-recovery and actionability analysis.
Figures~\ref{fig:microsoft-core} and~\ref{fig:microsoft-atq} summarize the same comparison visually, and Table~\ref{tab:microsoft-results} reports the Microsoft results. The Arch-Informed configuration recovers the largest portion of reference threats, with 0.83 raw coverage and 0.83 grounded coverage.
It also achieves the highest validity, traceability, actionability, and
attack-tree quality.

In addition, 25\% of the Arch-Informed validated findings are novel: these are
plausible design-time threats not listed in the Microsoft reference example but
judged relevant to the analyzed architecture by a security expert. Examples include integrity-check failure on Azure Storage and inadequate output
filtering from M3 to Azure Cognitive Services. These findings were not counted
as recovered reference threats, but they show that the framework can surface
additional architecture-specific review candidates beyond the provided threat
model.

Single-shot STRIDE generates the largest number of misuse cases, but its
grounded precision and actionability remain lower than Arch-Informed. This
suggests that structured threat categories alone do not provide the same level of
grounding or mitigation usefulness as the full pipeline. The PRD-only and
single-shot LLM baselines recover fewer reference threats and produce higher
noise, showing the value of architecture grounding and iterative refinement.

\textbf{Domain-level routing.}
To evaluate whether generated artifacts support enterprise security review, we
analyzed the domain tags assigned to the Microsoft Arch-Informed run. The
generated misuse and failure cases, as well as the downstream attack trees, were
routed across IAM, Network Security, Application Security, Data Security, Cloud
Security, DevSecOps, and GRC. Table~\ref{tab:microsoft-domain-routing}
summarizes the number of artifacts assigned to each domain. The distribution
shows that the framework does not produce a flat threat list; instead, it
organizes findings into reviewable security-domain workstreams. Data Security,
GRC, Network Security, Application Security, and Cloud Security receive the
largest number of routed artifacts, reflecting the data-flow, trust-boundary, and
cloud-service characteristics of the Microsoft IoT Edge architecture. Full domain-routing details for one Microsoft Arch-Informed run are provided in
Appendix~\ref{appendix:microsoft-details}.

\textbf{Candidate architecture redesign.}
The Microsoft Arch-Informed run produced a candidate revised architecture for
expert review; the original and revised diagrams are shown in
Appendix~\ref{appendix:microsoft-details}. Table~\ref{tab:microsoft-redesign-changes}
summarizes how the redesign maps validated findings to concrete controls,
including mutual TLS, RBAC-protected storage upload, data validation before
external cognitive-service processing, encrypted communication paths, telemetry
protection, and browser-side content-security controls.

\textbf{Representative token usage.}
The Microsoft Arch-Informed run required 43 LLM calls, 102{,}326 input tokens,
23{,}384 output tokens, and 125{,}710 total tokens. The largest token consumers
were attack-tree workers, misuse/failure-case generation, and confidence
scoring. Full token and cost details are provided in Appendix~\ref{appendix:microsoft-details}.

\begin{table}[!t]
\centering
\caption{Results on the Microsoft reference-labeled threat-modeling example.
Values are averaged over three trials per configuration. Coverage, grounded
coverage, grounded precision, and novelty are computed against the 12-threat
reference set. No-MC produces only downstream attack-tree artifacts, so
threat-scoped metrics are reported as N/A.}
\label{tab:microsoft-results}
\scriptsize
\setlength{\tabcolsep}{3pt}
\renewcommand{\arraystretch}{1.15}
\resizebox{\linewidth}{!}{
\begin{tabular}{lccccc}
\toprule
\textbf{Metric} &
\textbf{Arch-Informed} &
\textbf{No-MC} &
\textbf{PRD-only} &
\textbf{Single-shot LLM} &
\textbf{Single-shot STRIDE} \\
\midrule

\# Misuse cases generated & 13.3 & 0.0 & 4.7 & 5.7 & 17.3 \\
\# Attack trees generated & 9.0 & 6.7 & 2.3 & 3.0 & 6.0 \\

\midrule

Coverage (Raw) & 0.83 & N/A & 0.17 & 0.17 & 0.33 \\
Grounded Coverage & 0.83 & N/A & 0.08 & 0.17 & 0.11 \\
Grounded Precision & 0.63 & N/A & 0.10 & 0.30 & 0.23 \\
Validity & 1.00 & N/A & 0.55 & 0.50 & 0.67 \\
Noise & 0.00 & N/A & 0.45 & 0.50 & 0.33 \\
Traceability & 0.93 & N/A & 0.60 & 0.57 & 0.53 \\
Actionability & 0.97 & N/A & 0.65 & 0.57 & 0.33 \\
Novelty & 0.25 & N/A & 0.21 & 0.20 & 0.15 \\

\midrule

Attack-Tree Quality (1--5) & 5 & 3 & 4 & 2 & 2 \\

\bottomrule
\end{tabular}}
\end{table}

\begin{table}[!t]
\centering
\caption{Domain routing for the Microsoft Arch-Informed run. Counts indicate
artifacts tagged for each enterprise security domain.}
\label{tab:microsoft-domain-routing}
\scriptsize
\setlength{\tabcolsep}{2pt}
\renewcommand{\arraystretch}{1}
\resizebox{0.75\linewidth}{!}{
\begin{tabular}{lccc}
\toprule
\textbf{Domain} & \textbf{Misuse/Failure} & \textbf{Attack Trees} & \textbf{Total} \\
\midrule
IAM & 2 & 3 & 5 \\
Network Security & 4 & 4 & 8 \\
Application Security & 4 & 4 & 8 \\
Data Security & 5 & 6 & 11 \\
Cloud Security & 4 & 4 & 8 \\
DevSecOps & 1 & 2 & 3 \\
GRC & 6 & 6 & 12 \\
\bottomrule
\end{tabular}}
\end{table}

\begin{table*}[!t]
\centering
\caption{Candidate redesign changes for the Microsoft reference-labeled threat-modeling example: Arch-Informed run.}
\label{tab:microsoft-redesign-changes}
\scriptsize
\setlength{\tabcolsep}{4pt}
\renewcommand{\arraystretch}{1.08}
\begin{tabular}{p{3.2cm}p{4.2cm}p{4.4cm}p{3.2cm}}
\toprule
\textbf{Design change} &
\textbf{Diagram evidence} &
\textbf{Linked finding or weakness} &
\textbf{Security effect} \\
\midrule

Service-to-service authentication &
Mutual TLS is annotated on WebSocket flows among edge modules M1, M2, and M3. &
Missing independent authentication; unauthorized module-to-module access. &
Strengthens identity and trust between internal edge modules. \\

Storage upload access control &
RBAC is annotated on the feedback-frame upload path from M2 to Azure Storage. &
Unauthorized access to feedback uploads; weak access policy for cloud storage. &
Restricts storage writes to authorized modules and identities. \\

Input and request validation &
Data validation is annotated before the OCR/text-to-speech request to Azure Cognitive Services. &
Input manipulation; missing validation before external service processing. &
Reduces malformed or adversarial data sent to downstream services. \\

Encrypted data transmission &
Encrypted response and telemetry paths are annotated in the revised design. &
Telemetry data loss; MITM or disclosure risk on service communication paths. &
Improves confidentiality and integrity of data in transit. \\

Browser-side content security &
A Content Security Policy note is added for the browser-facing component. &
Frontend data-handling risk; inadequate output filtering or browser-side injection exposure. &
Reduces client-side injection and unsafe content-handling risk. \\

\bottomrule
\end{tabular}
\end{table*}

\subsection{Open-Ended Real-World Case Studies}

We next evaluate the framework on Berty and Gas Town, two 
open-source systems with different security-relevant design characteristics:
Berty represents secure peer-to-peer messaging, while Gas Town represents a
multi-agent workspace and workflow-management architecture. 
Because complete ground-truth threat sets are unavailable, we do not
report coverage, grounded coverage, grounded precision, or novelty. Instead, we report
validity, noise, traceability, actionability, redundancy, and attack-tree quality.

\begin{figure}[!t]
\centering
\includegraphics[width=0.75\linewidth]{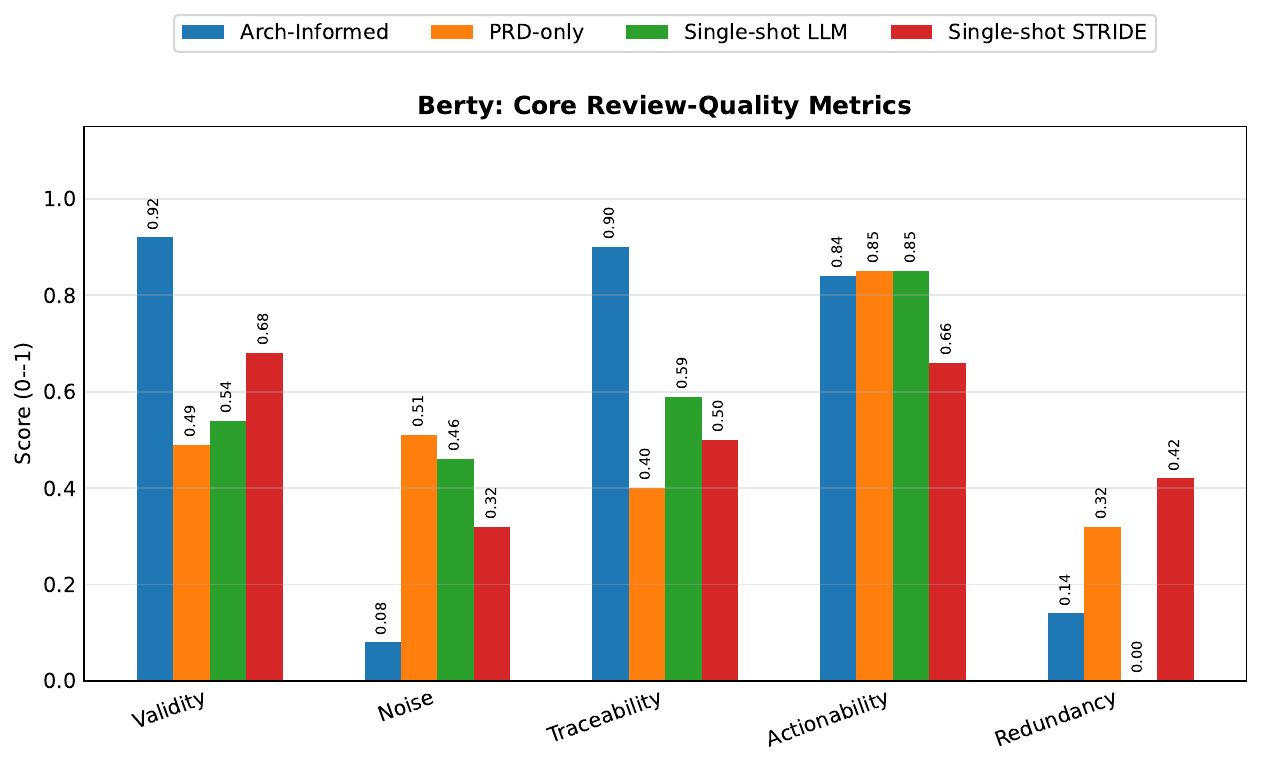}
\caption{Core review-quality metrics on Berty. No-MC is excluded because
threat-scoped metrics are not applicable when the misuse-case agent is removed.}
\label{fig:berty-core}
\end{figure}

\begin{figure}[!t]
\centering
\includegraphics[width=0.75\linewidth]{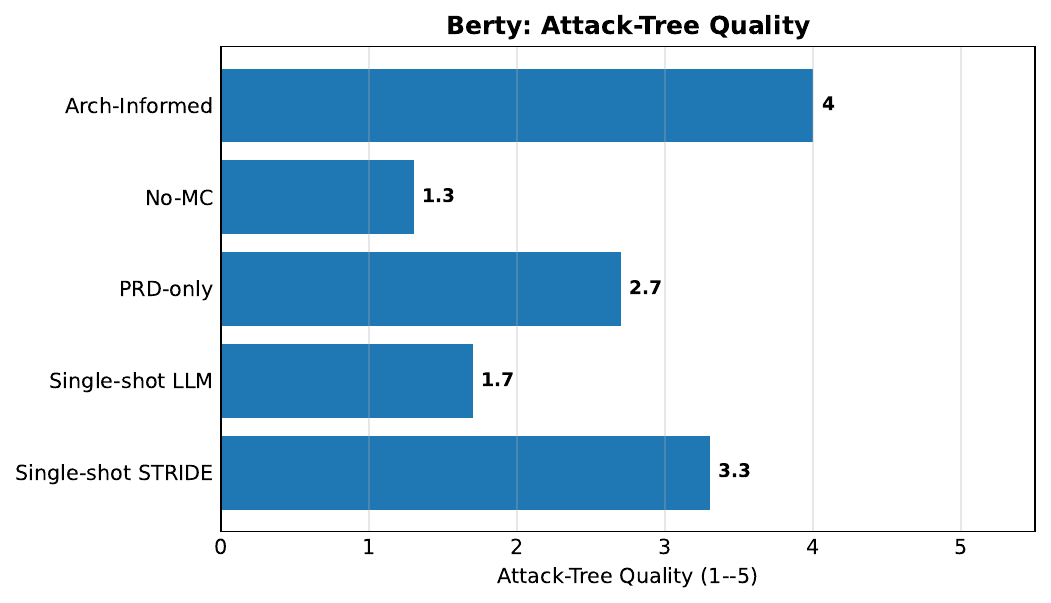}
\caption{Attack-tree quality on Berty. No-MC is included because it still
produces downstream attack-tree artifacts.}
\label{fig:berty-atq}
\end{figure}

\begin{figure}[!t]
\centering
\includegraphics[width=0.75\linewidth]{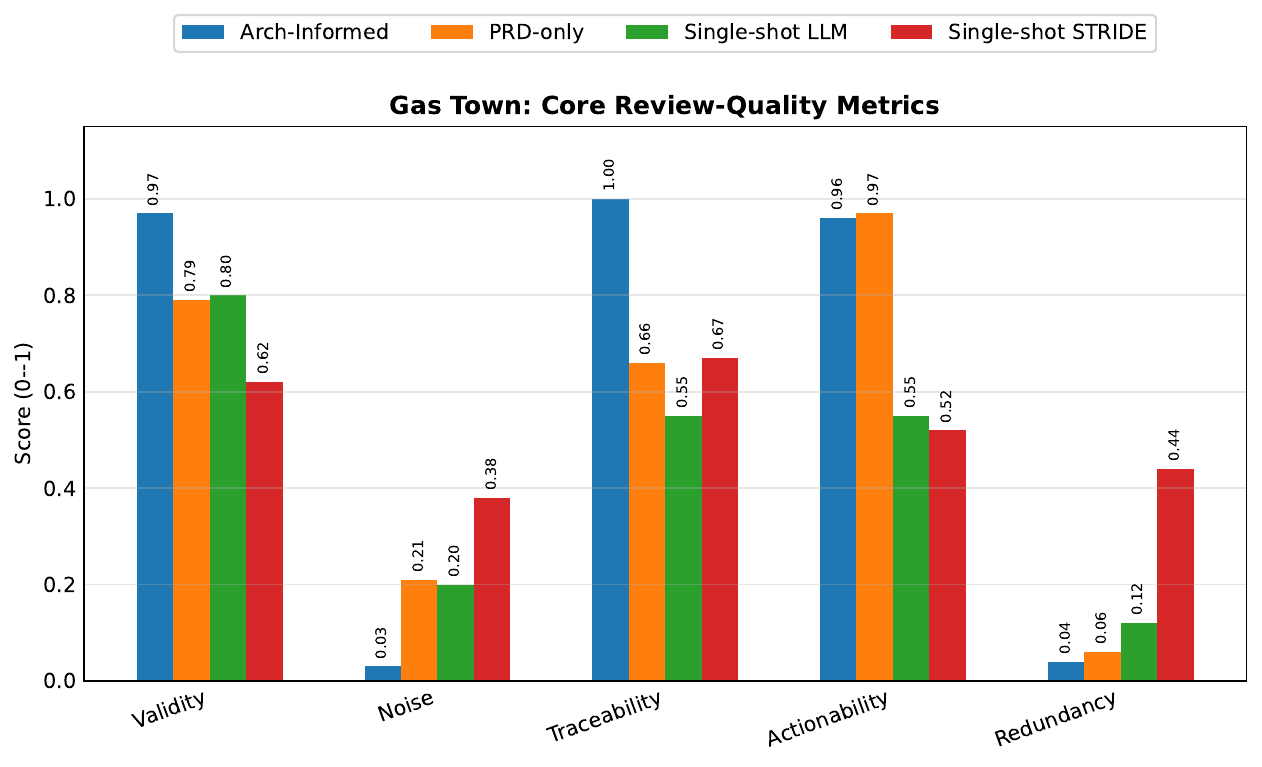}
\caption{Core review-quality metrics on Gas Town. No-MC is excluded because
threat-scoped metrics are not applicable when the misuse-case agent is removed.}
\label{fig:gastown-core}
\end{figure}

\begin{figure}[!t]
\centering
\includegraphics[width=0.75\linewidth]{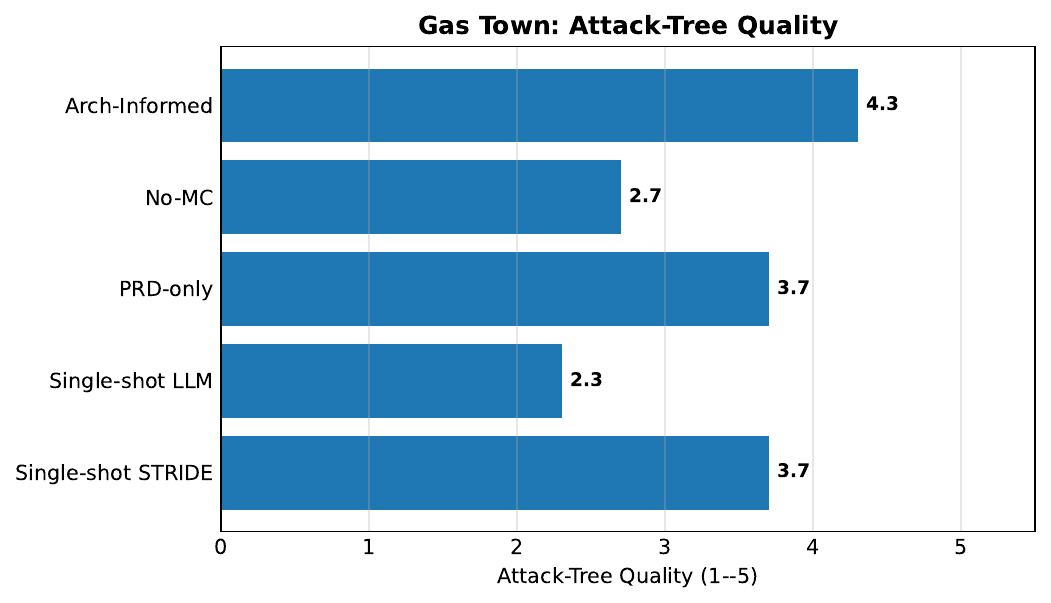}
\caption{Attack-tree quality on Gas Town. The Arch-Informed pipeline achieves
the highest score, while No-MC remains lower despite producing attack trees.}
\label{fig:gastown-atq}
\end{figure}

Table~\ref{tab:realworld-results} reports averaged results across three runs.
Figures~\ref{fig:berty-core}--\ref{fig:gastown-atq} visualize core metrics and attack-tree quality. The Arch-Informed
pipeline gives the strongest overall balance across both systems. STRIDE
generates a larger number of misuse cases, but it also produces the highest
redundancy, indicating overlapping findings created by taxonomy-driven
enumeration.

\begin{table*}[t]
\centering
\caption{Open-ended real-world results averaged over three runs. Coverage and
precision are omitted because no complete ground-truth threat set exists. N/A
indicates that threat-scoped metrics are not applicable for No-MC. \#MC: Misuse Cases Generated; \#AT: Attack Trees Generated.}
\label{tab:realworld-results}
\scriptsize
\setlength{\tabcolsep}{3pt}
\renewcommand{\arraystretch}{1.12}
\resizebox{\textwidth}{!}{
\begin{tabular}{llrrrrrrrr}
\toprule
\textbf{System} & \textbf{Config.} &
\textbf{\#MC} & \textbf{\#AT} &
\textbf{Valid.} & \textbf{Noise} & \textbf{Trace.} &
\textbf{Action.} & \textbf{Redun.} & \textbf{ATQ} \\
\midrule

Berty & Arch-Informed & 11.70 & 6.70 & 0.92 & 0.08 & 0.90 & 0.84 & 0.14 & 4.00 \\
Berty & No-MC & 0.00 & 4.70 & N/A & N/A & N/A & N/A & N/A & 1.30 \\
Berty & PRD-only & 6.30 & 5.00 & 0.49 & 0.51 & 0.40 & 0.85 & 0.32 & 2.70 \\
Berty & Single-shot LLM & 5.70 & 3.00 & 0.54 & 0.46 & 0.59 & 0.85 & 0.00 & 1.70 \\
Berty & Single-shot STRIDE & 19.30 & 4.70 & 0.99 & 0.01 & 0.50 & 0.98 & 0.42 & 3.30 \\

\midrule

Gas Town & Arch-Informed & 6.30 & 4.70 & 0.97 & 0.03 & 1.00 & 0.96 & 0.04 & 4.30 \\
Gas Town & No-MC & 0.00 & 6.70 & N/A & N/A & N/A & N/A & N/A & 2.70 \\
Gas Town & PRD-only & 4.70 & 2.30 & 0.79 & 0.21 & 0.66 & 0.97 & 0.06 & 3.70 \\
Gas Town & Single-shot LLM & 3.40 & 2.70 & 0.80 & 0.20 & 0.55 & 0.55 & 0.12 & 2.30 \\
Gas Town & Single-shot STRIDE & 23.40 & 4.00 & 0.65 & 0.35 & 0.67 & 0.04 & 0.44 & 3.70 \\

\bottomrule
\end{tabular}}
\end{table*}

For Berty, Arch-Informed achieves high validity (0.92), low noise (0.08),
strong traceability (0.90), and the highest attack-tree quality among the
pipeline configurations. STRIDE achieves high validity and actionability, but its
redundancy is substantially higher (0.42). This indicates that STRIDE produces
many overlapping findings even when individual findings are plausible.

For Gas Town, Arch-Informed is the most balanced configuration, reaching 0.97
validity, 0.03 noise, 1.00 traceability, 0.96 actionability, and 4.30 attack-tree
quality. STRIDE generates the largest number of misuse cases, but it also has
higher noise and redundancy, and lower actionability. This shows that structured
threat categories alone do not replace grounded misuse-case intermediates and
iterative validation.

Across all evaluation settings, the vulnerability categories most commonly surfaced by the Arch-Informed configuration fall into five design-level classes: (1) missing or weak service-to-service authentication across trust boundaries; (2) absent or insufficient input validation before external service calls; (3) overly broad access policies on sensitive data stores; (4) unprotected data flows across zone boundaries; and (5) missing audit, monitoring, or telemetry controls on high-sensitivity paths. These categories map naturally to the enterprise domain tags the framework assigns — IAM, Application Security, Data Security, Network Security, and DevSecOps respectively — serving as the framework's proxy for vulnerability classification given its design-level rather than implementation-level focus.

\subsection{Cross-Setting Analysis}

Across synthetic, reference-labeled, and open-ended settings, three patterns
emerge.

\textbf{Architecture grounding improves review quality.}
Arch-Informed achieves the strongest grounded precision on the synthetic cases
and the highest grounded coverage on the Microsoft reference-labeled case. On
Berty and Gas Town, where no complete ground truth is available, it maintains
high traceability, indicating that architecture context helps link findings to
components, data flows, and trust boundaries.

\textbf{Misuse-case intermediates improve reliability.}
No-MC can increase apparent coverage, as in OLB, but this comes with lower
grounded precision. In the Microsoft and open-ended cases, No-MC does not
support threat-scoped metrics because removing the misuse-case agent removes the
validated threat unit used for evaluation. This supports the role of misuse
cases as grounded intermediate artifacts.

\textbf{Single-shot and STRIDE baselines are less balanced.}
Single-shot LLM outputs show weaker grounded precision, traceability, and
attack-tree quality. STRIDE improves structure and breadth, but often increases
redundancy because multiple findings describe the same weakness from different
STRIDE perspectives. Thus, taxonomy-driven enumeration is useful as a baseline,
but it does not replace architecture-grounded misuse cases, validation, and
refinement.

Overall, Arch-Informed does not merely generate more findings; it produces more
reviewable findings. The synthetic cases show controlled recovery of known
flaws, the Microsoft case shows recovery against a reference threat model, and
the Berty and Gas Town cases show that the same pipeline remains useful in
open-ended architectures.

\section{Discussion}

\subsection{Answering the Research Questions}

\textbf{RQ1: Can LLM agents identify design-time security weaknesses from PRDs
and architecture diagrams?}
The results show that, in our evaluated settings, LLM agents can identify meaningful design-time security
weaknesses when their reasoning is grounded in both requirements and
architecture. In the controlled synthetic cases, the Arch-Informed pipeline
recovers a large portion of known design flaws. In the Microsoft
reference-labeled case, it recovers the largest portion of reference threats and
also surfaces plausible additional review candidates. In the open-ended Berty
and Gas Town cases, the same configuration maintains high validity,
traceability, and actionability without relying on complete ground truth.

\textbf{RQ2: Do misuse-case intermediates and validation improve
grounding, traceability, and noise reduction?}
Yes. Across settings, removing misuse-case intermediates weakens the connection
between generated findings and downstream reasoning artifacts. No-MC can
increase apparent coverage in some controlled cases, but this comes with lower
grounded precision and weaker traceability. The full pipeline uses misuse cases
as stable intermediate artifacts that connect requirements, components, attack
paths, mitigations, and domain tags.

\textbf{RQ3: Can generated findings support mitigation planning, enterprise-domain
routing, and secure architecture redesign?}
The results show that generated findings can support downstream review tasks,
including mitigation planning, enterprise-domain routing, and candidate
architecture redesign. In the Microsoft case, the Arch-Informed run routes
findings across IAM, Network Security, Application Security, Data Security,
Cloud Security, DevSecOps, and GRC, and produces a revised architecture with
concrete design changes such as mutual TLS, RBAC-protected storage uploads,
data validation, encrypted communication paths, telemetry protection, and
browser-side content-security controls. These outputs should be interpreted as
structured review aids for security architects, not as automatically approved
designs.

\subsection{Why Misuse Cases Matter}

A key finding is that misuse cases act as grounding anchors between design
artifacts and downstream security reasoning. Without this intermediate layer,
the pipeline can still generate attack trees or STRIDE-style threat lists, but the
outputs are less traceable and more redundant. Misuse cases provide a stable
unit of analysis: each case links an actor, affected component, violated security
property, precondition, impact, and supporting evidence. This structure helps
the pipeline validate findings, refine weak artifacts, and generate targeted
mitigations.

\subsection{Controlled vs. Open-Ended Evaluation}

The synthetic results should be interpreted as controlled detection results. The
synthetic PRD--architecture pairs were intentionally seeded with design flaws,
so higher recovery scores are expected. Their value is that they allow controlled
measurement of coverage, grounded precision, and mitigation quality. The
Microsoft case provides a stronger reference-labeled validation setting because
the threats and mitigations were not injected by us. Berty and Gas Town provide
open-ended evidence that the framework remains useful when complete ground
truth is unavailable.

\subsection{Practical Use in Enterprise Security Review}

The framework is designed to support, not replace, human security review. Its
main value is in accelerating early-stage design analysis by producing grounded
findings, attack paths, mitigations, and domain labels before implementation
begins. The domain tags help route findings to IAM, network security,
application security, data security, cloud security, DevSecOps, and GRC teams.
This makes the output more aligned with enterprise review workflows than a
flat list of model-generated threats.

\textbf{Evaluator judgment.}
Metrics such as validity, actionability, novelty, and attack-tree quality require
human judgment. To reduce subjectivity, we use explicit scoring criteria and
evaluate outputs across multiple runs. The evaluator was an enterprise security expert, and all outputs were reviewed with a fixed rubric.

\section{Limitations and Threats to Validity}

\textbf{Synthetic evaluation.}
The synthetic systems are intentionally flawed and therefore may inflate recovery
metrics. We use them as controlled stress tests rather than as evidence of broad
generalization. To reduce this risk, we also evaluate the framework on a
reference-labeled Microsoft threat model and two open-ended
systems.

\textbf{Ground-truth completeness.}
Complete ground-truth threat sets are difficult to define for open-ended
real-world architectures. For Berty and Gas Town, we therefore do not report
coverage or precision. Instead, we evaluate review-quality metrics such as
validity, noise, traceability, actionability, redundancy, and attack-tree quality.

\textbf{Scope of design artifacts.}
The current evaluation focuses on PRDs and architecture diagrams. The framework
does not yet extract architecture directly from codebases or runtime telemetry.
Future work can extend the input stack to code repositories, API specifications,
deployment manifests, and operational traces.

\textbf{Adversarial Input Artifacts.}
A threat the current framework does not address is adversarial manipulation of input artifacts. A PRD or architecture diagram containing carefully crafted content could attempt to influence agent outputs through prompt injection. For example, embedding instructions in diagram labels or requirement text that redirect agent reasoning. The framework currently treats all input content as trusted. Defenses such as input sanitization, agent-level output verification against the source artifacts, and sandboxed agent execution are not yet implemented and represent a meaningful attack surface for enterprise deployments.

\textbf{Architecture redesign.}
The revised Microsoft architecture is evaluated as a candidate design artifact
for expert review, not as a deployed implementation. The diagram demonstrates
that the framework can translate validated findings into concrete design-change
recommendations, but it does not prove that the redesigned system is complete,
optimal, or implementation-ready.

\textbf{Human approval.}
The revised architecture diagrams generated by the framework are design
recommendations, not approved final architectures. Human architects and security
reviewers must validate feasibility, cost, compliance impact, and implementation
priority before adoption.

\textbf{Evaluator subjectivity.} Metrics requiring human judgment such as validity, actionability, novelty, and attack-tree quality, were assessed by a single enterprise security expert. Inter-rater reliability was not formally measured. Future work should engage multiple independent evaluators and report agreement scores such as Cohen's $\kappa$ to strengthen the validity of these assessments.

\textbf{Refinement loop isolation.} The current ablations do not isolate the iterative refinement loop independently from the misuse-case agent. The No-MC ablation removes both the misuse-case intermediates and their role in guiding refinement, so improvements in the full Arch-Informed pipeline reflect the combined effect of architecture grounding, misuse-case intermediates, and iterative validation. A no-refinement ablation — retaining misuse cases but disabling the validation loop — would isolate the contribution of refinement alone and is planned for future work.
\section{Conclusion}

This paper presented a grounded multi-agent LLM framework for design-time
vulnerability discovery and secure architecture redesign from PRDs and
architecture diagrams. The framework parses architecture artifacts into graph
representations, generates misuse and failure cases, constructs attack trees,
validates and refines outputs, identifies governance gaps, recommends
mitigations, assigns enterprise security-domain tags, and produces candidate revised architecture recommendations.

Across controlled synthetic systems, a Microsoft reference-labeled threat model,
and open-ended Berty and Gas Town case studies, the results show that
architecture-informed, misuse-driven reasoning improves grounding,
traceability, actionability, and attack-tree quality compared with single-shot
and ablation baselines. The results also show that larger threat lists are not
necessarily better: STRIDE-style enumeration can increase redundancy, while
misuse-case intermediates and validation help produce more reviewable findings.

The framework is designed to support human security reviewers and architects,
not replace them. Its revised architectures are structured recommendations that
require expert approval before adoption. Future work will evaluate the approach
on additional real-world systems, larger enterprise artifacts, multiple LLM
backends, and richer inputs such as API specifications, deployment manifests,
and code-derived architecture views.
\newpage

\raggedright
\bibliographystyle{iclr2025_conference}
\bibliography{contents/references}

\newpage
\appendix

\begin{figure}[t]
\centering
\includegraphics[width=\linewidth]{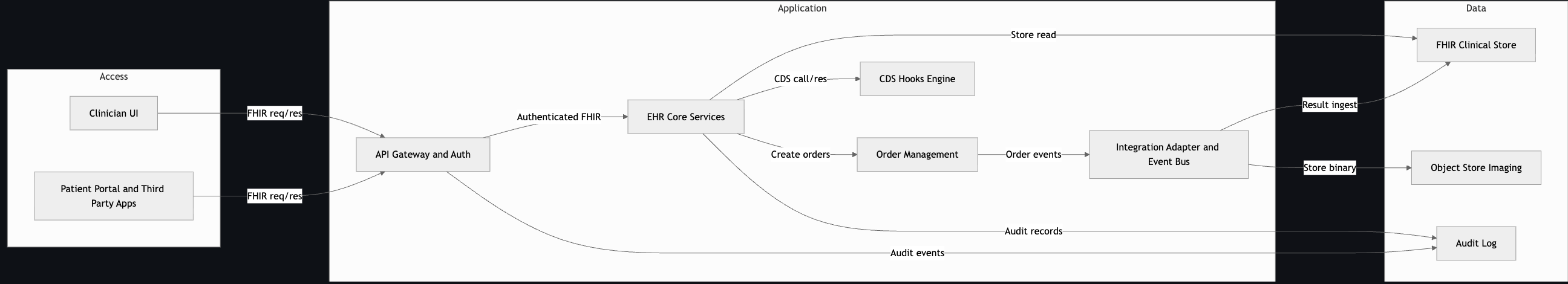}
\caption{Synthetic Electronic Health System (EHS) architecture.}
\label{fig:eha}
\end{figure}

\begin{figure}[t]
\centering
\includegraphics[width=\linewidth]{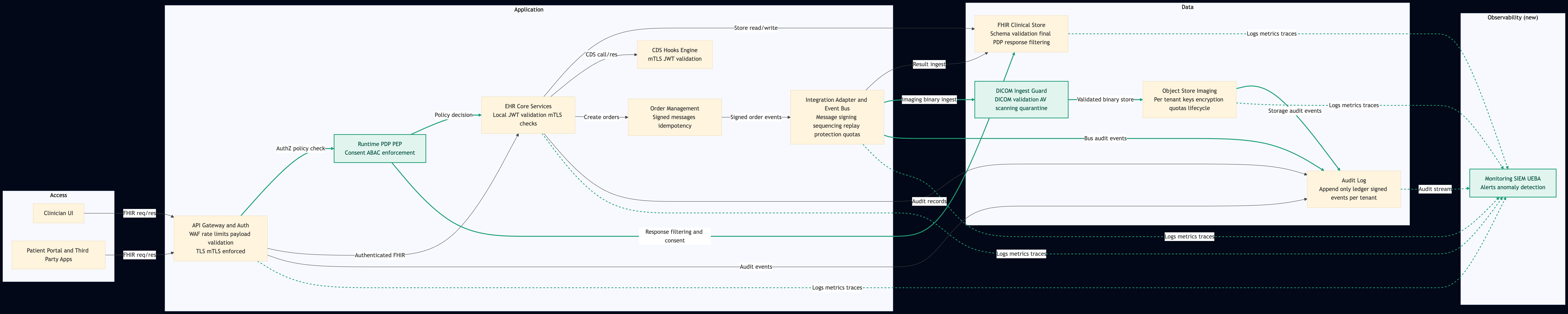}
\caption{Candidate revised Electronic Health System (EHS) architecture.}
\label{fig:eha-new}
\end{figure}
\section{Microsoft Redesign and Domain-Routing Details}
\label{appendix:microsoft-details}

\begin{figure}[t]
\centering
\includegraphics[width=\linewidth]{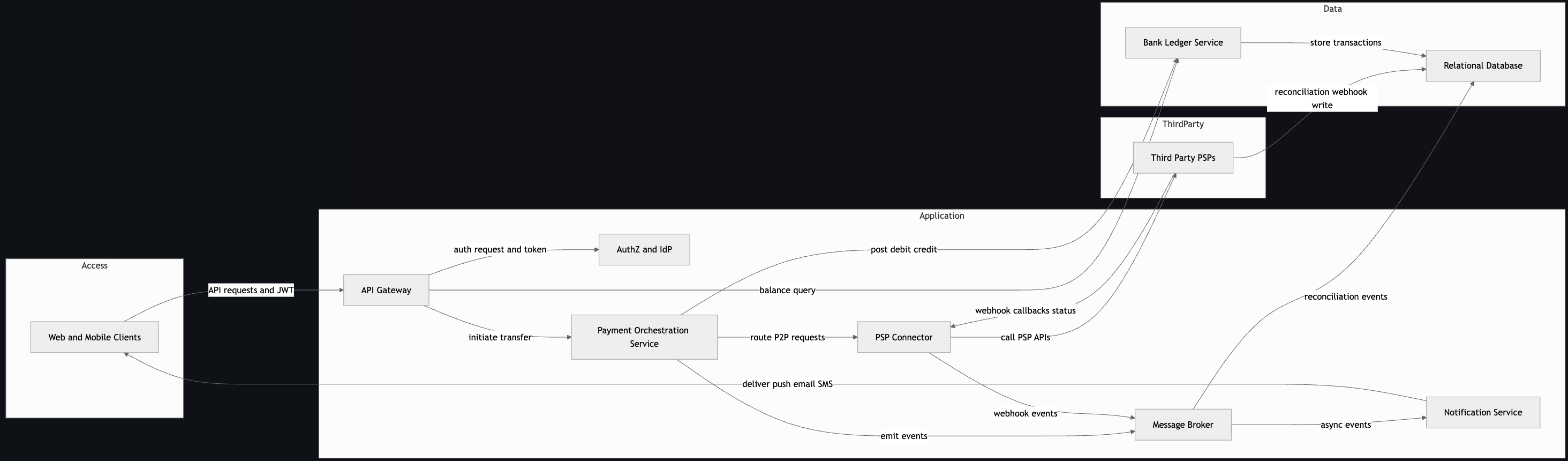}
\caption{Synthetic Online Banking System (OLB) architecture.}
\label{fig:olb}
\end{figure}

This appendix provides supporting artifacts for the Microsoft
reference-labeled Arch-Informed run: the original and candidate revised
architecture, detailed domain-routing assignments, and representative token
usage. Domain tags use the following abbreviations: IAM = Identity and Access
Management, NET = Network Security, APP = Application Security, DATA = Data
Security, CLOUD = Cloud Security, DEV = DevSecOps, and GRC = Governance, Risk,
and Compliance.

\begin{figure*}[t]
\centering
\begin{minipage}{0.49\textwidth}
\centering
\includegraphics[width=\linewidth]{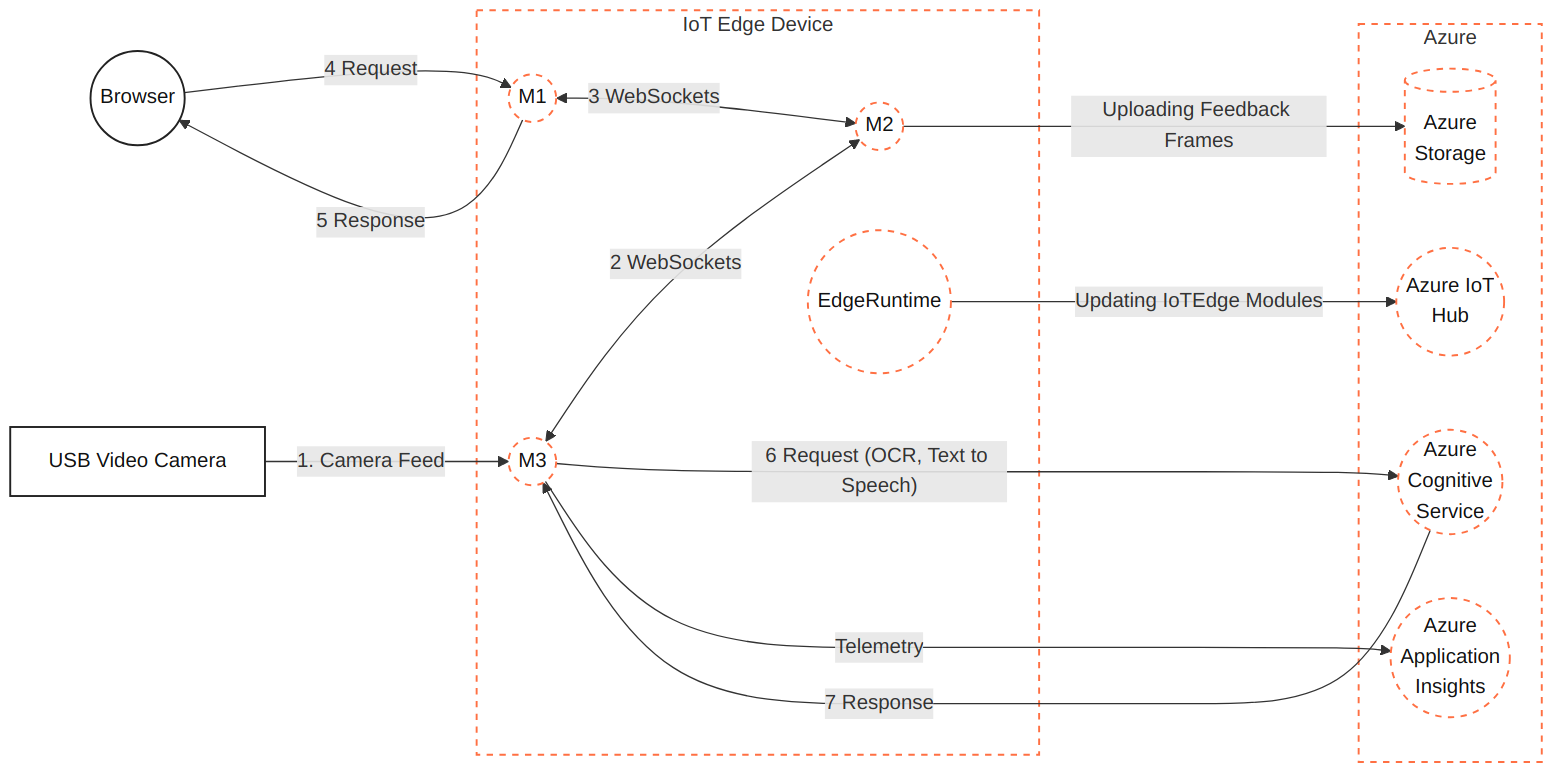}
\textbf{(a) Original architecture}
\end{minipage}
\hfill
\begin{minipage}{0.49\textwidth}
\centering
\includegraphics[width=\linewidth]{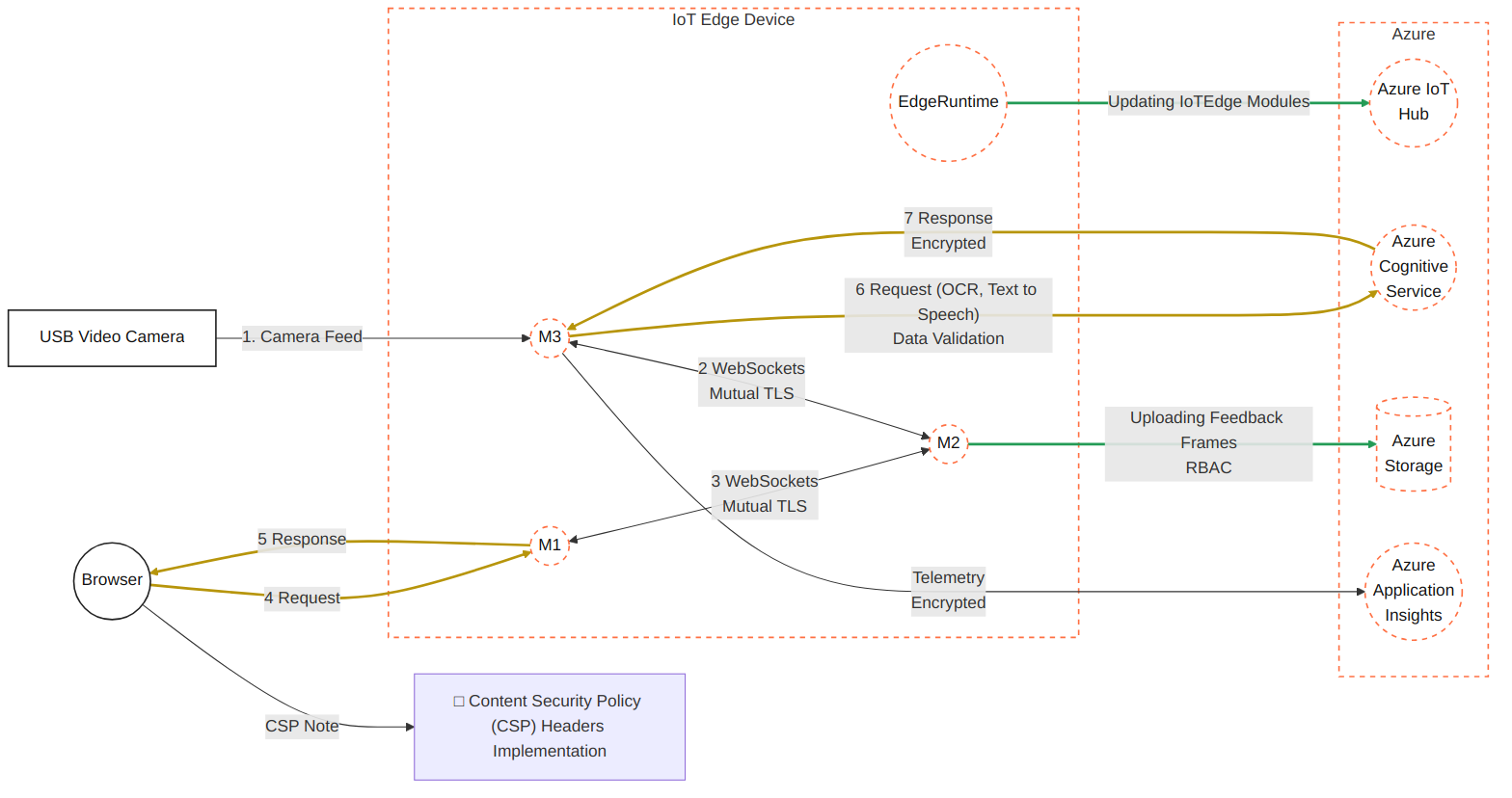}
\textbf{(b) Candidate revised architecture}
\end{minipage}
\caption{Original and candidate revised architecture for the Microsoft
reference-labeled threat-modeling case. The revised architecture adds
design-time controls derived from Arch-Informed findings and mitigation
recommendations, including mutual TLS, RBAC-protected storage upload, data
validation, encrypted communication paths, telemetry protection, and
browser-side content-security controls.}
\label{fig:microsoft-redesign}
\end{figure*}

\begin{table*}[t]
\centering
\caption{Detailed domain routing for the Microsoft Arch-Informed run.}
\label{tab:microsoft-domain-routing-details}
\scriptsize
\setlength{\tabcolsep}{4pt}
\renewcommand{\arraystretch}{1.03}
\begin{tabular}{p{1.15cm}p{9.6cm}p{4.2cm}}
\toprule
\textbf{Type} & \textbf{Artifact} & \textbf{Domain tags} \\
\midrule
MC & No Rate Limiting on IoTEdge Modules & NET, GRC \\
MC & Missing Independent Authentication & IAM, NET, APP, GRC \\
MC & Lack of Fine-Grained Authorization & IAM, DATA, GRC \\
MC & Inadequate Output Filtering $(M3 \rightarrow \mathrm{Cognitive})$ & NET, DATA \\
MC & Missing Input Validation in Modules & APP, GRC \\
MC & Unauthorized Access to Cognitive Transfers & NET, DATA, CLOUD, GRC \\
MC & Integrity Check Failure on AzureStorage & DATA, CLOUD \\
MC & Unauthorized Access via Camera Feed & APP, DATA, GRC \\
MC & EdgeRuntime Deployment Failure & CLOUD, DEV \\
MC & Input Manipulation via Browser & APP \\
MC & Telemetry Data Loss to App Insights & CLOUD \\
\midrule
AT & 4.1 DoS via Uncontrolled Public Access & NET, GRC \\
AT & 4.2 Unauthorized Access to AzureStorage (Trust Boundaries) & IAM, NET, APP, GRC \\
AT & 4.3 Inadequate Authorization $\rightarrow$ Unauthorized Uploads & IAM, DATA, GRC \\
AT & 4.4 Data Integrity via Unvalidated Inputs & APP, DATA, GRC \\
AT & 4.5 MITM on $M3 \rightarrow \mathrm{AzureCognitive}$ & NET, DATA, CLOUD, GRC \\
AT & 4.6 Inadequate Output Filtering $M3 \rightarrow \mathrm{Cognitive}$ & NET, DATA \\
AT & 4.7 Unauthorized Access to Camera Feed & APP, DATA, GRC \\
AT & 4.8 Data Loss in Telemetry Channel & CLOUD, DEV \\
AT & 4.9 Compromise Azure Key Vault / Secrets & IAM, APP, DATA, CLOUD \\
AT & 4.10 EdgeRuntime Deployment Failure & CLOUD, DEV \\
\bottomrule
\end{tabular}
\end{table*}

\begin{table}[t]
\centering
\caption{Representative token usage for the Microsoft Arch-Informed run.}
\label{tab:microsoft-token-usage}
\scriptsize
\setlength{\tabcolsep}{5pt}
\renewcommand{\arraystretch}{1.03}
\begin{tabular}{lrrrrr}
\toprule
\textbf{Stage} &
\textbf{Calls} &
\textbf{Input Tok.} &
\textbf{Output Tok.} &
\textbf{Total Tok.} &
\textbf{Share} \\
\midrule
Misuse / failure-case generation & 8 & 26{,}833 & 5{,}655 & 32{,}488 & 25.8\% \\
Attack-tree planner & 3 & 5{,}392 & 515 & 5{,}907 & 4.7\% \\
Attack-tree workers & 17 & 27{,}491 & 6{,}914 & 34{,}405 & 27.4\% \\
Confidence scoring / reasoning & 8 & 23{,}317 & 6{,}907 & 30{,}224 & 24.0\% \\
Attack-tree validation & 3 & 13{,}772 & 1{,}242 & 15{,}014 & 11.9\% \\
Governance review & 1 & 2{,}625 & 901 & 3{,}526 & 2.8\% \\
Recommendations & 2 & 2{,}889 & 1{,}240 & 4{,}129 & 3.3\% \\
Other / unassigned overhead & 1 & 7 & 10 & 17 & $<$0.1\% \\
\midrule
\textbf{Total} & \textbf{43} & \textbf{102{,}326} & \textbf{23{,}384} &
\textbf{125{,}710} & \textbf{100\%} \\
\bottomrule
\end{tabular}
\end{table}

\section{Synthetic Architectures}
\label{appendix:synthetic-arch}

The synthetic cases use editable architecture diagrams so that the framework can
generate candidate redesigned diagrams in addition to threat findings, attack
trees, governance observations, and mitigation recommendations. The synthetic
systems include explicit gaps, such as missing authentication, and implicit
omissions, such as absent validation or weak inter-service trust. The OLB and
EHS ground-truth sets each contain 12 labeled threats, split between misuse
cases and workflow or service-level threats.

The Online Banking System (OLB) is a synthetic retail banking platform where
web and mobile clients reach payment and account services through an API
gateway, backed by a bank ledger and relational database, with peer-to-peer
payments handled by external providers. Injected weaknesses include missing
risk screening before ledger updates, unverified PSP webhooks, and missing
resource-level authorization on account and transaction data.

The Electronic Health System (EHS) is a synthetic clinical platform where
clinician and patient-facing applications reach record, ordering, and
decision-support services through an API gateway, backed by a FHIR clinical
store, imaging store, and audit log. Injected weaknesses include a read path
without consent checking, unvalidated stored data, and missing service-to-service
authentication. Figure~\ref{fig:eha-new} shows the candidate redesigned EHS
architecture generated after mitigation recommendations.

\section{Cost Estimation}
\label{costestimation}
Table \ref{tab:microsoft-token-usage} shows the representative token usage for different agents for Microsoft Arch-Informed run.
Cost is estimated from input and output token counts using:
\[
\mathrm{Cost}
=
\left(\frac{T_{in}}{10^6}\right)P_{in}
+
\left(\frac{T_{out}}{10^6}\right)P_{out},
\]
where $T_{in}$ and $T_{out}$ are input and output tokens, and $P_{in}$ and
$P_{out}$ are the provider prices per million input and output tokens at the
time of the experiment.

For the Microsoft Arch-Informed run, this gives:
\[
\mathrm{Cost}
=
0.102326P_{in} + 0.023384P_{out}.
\]

\end{document}